\documentclass[12pt]{iopart}

\usepackage{iopams}
\usepackage{cite}
\usepackage{txfonts}
\usepackage{mathptmx}
\usepackage{graphicx}
\usepackage{float}
\usepackage{subfigure}
\begin{document}

\title[Nonlocal symmetries related to B\"acklund transformation and their applications]
{Nonlocal symmetries related to B\"acklund transformation and
their applications}

\author{S Y Lou$^{1,3}$, Xiaorui Hu$^{2,1}$ and Yong Chen$^{2,1}$}

\address{$^{1}$Faculty of Science, Ningbo University, Ningbo,
               Zhejiang 315211, People's Republic of China}
\address{$^{2}$Shanghai Key Laboratory of Trustworthy Computing, East China
                Normal University, Shanghai 200062, China}
\address{$^{3}$Department of Physics, Shanghai Jiao Tong University,
               Shanghai 200240, China}
\ead{lousenyue@nbu.edu.cn, ychen@sei.ecnu.edu.cn}

\begin{abstract}
Starting from  nonlocal symmetries related to  B\"acklund
transformation (BT), many interesting results can be obtained.
Taking the well known potential KdV (pKdV) equation as an example,
a new type of nonlocal symmetry in elegant and compact form which
comes from BT is presented and used to make researches in the
following three subjects: two sets of negative pKdV hierarchies and
their corresponding bilinear forms are constructed; the nonlocal
symmetry is localized by introduction of suitable and simple
auxiliary dependent variables to generate new solutions from old ones and
to consider some novel group invariant solutions; some other models
both in finite dimensions and infinite dimensions are generated  by
comprising the original BT and evolution under new nonlocal
symmetry. The finite-dimensional models are completely integrable in
Liouville sense, which are shown equivalent to the results given
through the nonlinearization method for Lax pair.
\end{abstract}
%\submitto{\JPA}
%\maketitle
\section{Introduction}
With the development of integrable systems and solion theory,
symmetries \cite{1,2,3} play the more and more important role in
nonlinear mathematical physics. Thanks to the classical or
nonclassical Lie group method, Lie point symmetries of a
differential system can be obtained, from which one can transform
given solutions to new ones via finite transformation and construct
group invariant solutions by similarity reductions. However, little
importance is attached to the existence and applications of nonlocal
symmetries \cite{2,3}. Firstly, seeking for nonlocal symmetries in
itself is a difficult work to perform. One of our authors (Lou) has
made some efforts to get infinite many nonlocal symmetries by
inverse recursion operators \cite{4,5} the conformal invariant form
(Schwartz form) \cite{6} and Darboux transformation \cite{7,8}.
Moreover, it appears that the nonlocal symmetries are rarely used to
construct explicit solutions since the finite symmetry
transformations and similarity reductions can not be directly
calculated under the nonlocal symmetries. Naturally, it is necessary
to inquire as to whether nonlocal symmetries can be transformed to
local ones. The introduction of potential \cite{3} and
pseudopotential type symmetries \cite{9,10,11} which possesses close
prolongation extends the applicability of symmetry methods to obtain
solutions of differential equations (DEs). In that context, the
original given equation(s) can be embedded in some prolonged
systems. Hence, these nonlocal symmetries with close prolongation
are anticipated \cite{12,13,14}.

On the other hand, to find new integrable models is another
important application of  symmetry study. A systematic approach have
been developed by Cao \cite{15,16,17} to find finite-dimensional
integrable systems by the nonlinearization of Lax pair under certain
constraints between potentials and eigenfunctions. Especially in the
study of (1+1)-dimensional soliton equations, various new kinds of
confocal involutive systems are constructed by the approach of
nonlinearization of eigenvalue problems or constrained flows
\cite{18,19}. It has also been pointed that by restricting a
symmetry constraint to the Lax pair of soliton equation, one can not
only obtain the lower dimensional integrable models from higher
ones, but also embed the lower ones into higher dimensional
integrable models \cite{6,8,20}.  Here, alternatively, we are
inspired to act the given nonlocal symmetry on the B\"acklund
transformation (BT) instead of Lax pair to generate some  other new
systems via symmetry constraint method. The related work may be
adventurous but full of enormous interest.

In this paper, taking the well known potential KdV equation (pKdV)
for a special example, we will study the nonlocal symmetry defined
by BT. Since the BT reveals a finite transformation between two
exact solutions of DEs, it must hint some symmetry. For pKdV
equation, a new class of nonlocal symmetries are derived from its
BT, which may give more interesting applications than those nonlocal
symmetries only including potentials and pseudopotentials. The
prolongation of the new nonlocal symmetries are found close after
extending pKdV equation to an auxiliary system with four dependent
variables. The finite symmetry transformation and similarity
reductions are computed to give exact solutions of KdV equation.
What we want to mention is the process can  once lead to two exact
solutions from one given result due to the B\"acklund
transformation. Moreover, for the pKdV equation, some other models
both in finite dimensions and infinite dimensions are obtained.  The
finite-dimensional systems obtained here are found equivalent to the
results given by Cao \cite{16}, which have been verified completely
integrable in Liouville sense. This discovery confirms that these
obtained infinite-dimensional models should have many nice
integrable properties, which needs our further study.

The paper is organized as follows. In section II, we present a
detailed description about the new nonlocal symmetry with BT of pKdV equation.
Two kinds of flow equations
corresponding to the given nonlocal symmetry, i.e. the negative pKdV
hierarchies, are obtained and their corresponding bilinear forms are
also given out. In section III, we extend the nonlocal symmetry to
be equivalent to a Lie point symmetry of a auxiliary  prolonged
system admitting pKdV equation and its BT. Then the finite symmetry
transformation and similarity reductions are made to produce exact
solutions of pKdV and then KdV equation. Section IV is devoted to constructing
various integrable systems by means of symmetry constraint method.
Conclusions and discussions are given in Section V.

\section{Nonlocal symmetries and flow equations related to BT}

\subsection{BT for the pKdV equation}

The well-known KdV equation reads
\begin{eqnarray}
\omega_t+\omega_{xxx}-6\omega\omega_x=0,\label{kdv}
\end{eqnarray}
where subscripts $x$ and $t$ denote partial differentiation. For
convenience to deal analytically with a potential function $u$,
introduced by setting $\omega=u_x$, it follows from equation (1)
that $u$ would satisfy the equation
\begin{eqnarray}
u_t+u_{xxx}-3u^2_x=0, \label{pkdv}
\end{eqnarray}
which is called potential KdV (pKdV) equation.

For equation (2), there exists the following BT \cite{21}
\begin{eqnarray}
u_x+u_{1,x}=-2\lambda+\frac{(u-u_1)^2}{2}, \label{BTx}\\
u_t+u_{1,t}=2u^2_{x}+2u^2_{1,x}+2u_xu_{1,x}-(u-u_1)(u_{xx}-u_{1,xx})\label{BTt}
\end{eqnarray}
with $\lambda$ being arbitrary parameter.

Equations (3) and (4) show that if $u$ is a solution of equation
(2), so is $u_1$, that is to say, they represent a finite symmetry
transformation between two exact solutions of equation (2).

On the other hand, equations (3) and (4) can also be viewed as a
nonlinear Lax pair of equation (2). For
\begin{eqnarray}
u_{1,x}=-u_x-2\lambda+\frac{(u-u_1)^2}{2},\label{5}\\
u_{1,t}=-u_t+2u^2_{x}+2u^2_{1,x}+2u_xu_{1,x}-(u-u_1)(u_{xx}-u_{1,xx}),
\label{6}
\end{eqnarray}
its compatibility condition $u_{1,xt}=u_{1,tx}$ is  exactly equation
(2). In fact, both equation (3) and equation (4) hint that they are
all Riccati type equations about $u$ or $u_1$, which can be
linearized by the well known Cole-Hopf transformation
\begin{eqnarray}
u=-2\frac{\psi_x}{\psi}, \qquad {\rm{or}} \qquad
u_1=-2\frac{\psi_{1,x}}{\psi_1}. \label{7}
\end{eqnarray}

Moreover, by virtue of the dependent variables transformation
(\ref{7}), one can convert equation (2) into the following  bilinear
form
\begin{eqnarray}
(D^4_x+D_xD_t)\psi\cdot\psi=0, \label{8}
\end{eqnarray}
meanwhile it leads equations (3) and (4) to
\begin{eqnarray}
(D^2_x-\lambda) \psi\cdot \psi_1= 0,\label{9}\\
(D_t+D^3_x+3\lambda D_x)\psi\cdot\psi_1=0, \label{10}
\end{eqnarray}
where the Hirota's bilinear operator $D^m_xD^n_t$ is defined by
\[
D^m_xD^n_t a\cdot b=\left.\left(\frac{\partial}{\partial
x}-\frac{\partial}{\partial x'}\right)^m
\left(\frac{\partial}{\partial t}-\frac{\partial}{\partial
t'}\right)^n a(x,t)b(x',t')\right|_{x'=x,t'=t}.
\]

\subsection{The nonlocal symmetry from B\"acklund transformation}

For equation (\ref{pkdv}) with its BT (\ref{BTx}) and (\ref{BTt}),
considering the invariant property under
\[
\lambda \rightarrow \lambda+\epsilon \delta, \qquad u \rightarrow
u+\epsilon \sigma,\qquad u_1 \rightarrow u_1+\epsilon \sigma',
\nonumber
\]
we may find substantial possible  nonlocal symmetries and  a special
case is presented and studied as follows.\\

\textbf{\emph{Proposition 1.}} The pKdV equation (2) has a new type
of nonlocal symmetry given by
\begin{eqnarray}
\sigma=\exp(\int {u-u_1}\rmd x), \label{12}
\end{eqnarray}
where $u$ and $u_1$ satisfy BT (3) and (4). That means $\sigma$
given by (\ref{12}) satisfies the following symmetry equation
\begin{eqnarray}
\sigma_t+\sigma_{xxx}-6u_x\sigma_x=0. \label{11}
\end{eqnarray}
\emph{Proof:} By direct calculation.

On the other hand, we let the bilinear pKdV equation (\ref{8}) be
invariant under the transformation $\psi\rightarrow\psi+\epsilon
\sigma_{\psi}$, which produces the corresponding symmetry equation
\begin{eqnarray}
(D^4_x+D_xD_t)\sigma_{\psi}\cdot\psi=0. \label{13}
\end{eqnarray}
The Cole-Hopf transformation $u=-2\frac{\psi_x}{\psi}$ between
equation (2) and its bilinear equation (\ref{8}) determines a
symmetry transformation for $\sigma$ and $\sigma_{\psi}$, saying
\begin{eqnarray}
\sigma=\frac{2\psi_x\sigma_{\psi}}{\psi^2}-\frac{2\sigma_{\psi,x}}{\psi}.
\label{14}
\end{eqnarray}
Taking equations (\ref{12}) and (\ref{7}) into equation (\ref{14}),
we obtain a class of  nonlocal symmetry for equation (\ref{8})
\begin{eqnarray}
\sigma_{\psi}=-\frac{\psi}{2}\int \frac{\psi^2_1}{\psi^2}dx.
\label{15}
\end{eqnarray}
Correspondingly, it gives the following proposition for equation
(\ref{8}).\\

 \textbf{\emph{Proposition 2.}} The bilinear pKdV equation
(\ref{8}) has the nonlocal symmetry expressed by (\ref{15}), where
$\psi$ and $\psi_1$ satisfy
bilinear BT (\ref{9}) and (\ref{10}). \\
\emph{Proof:} One can directly check that $\sigma_{\psi}$ given by
(\ref{15}) satisfies symmetry equation (\ref{13}) under the
consideration (\ref{9}) and (\ref{10}).

\subsection{Two sets of negative pKdV hierarchies}

The existence of infinitely many symmetries leads to the  the
existence of integrable hierarchies and with the help of  infinitely
many nonlocal symmetries, one can extend the original system to its
negative hierarchies \cite{22,23}. Here, starting from the nonlocal
symmetry (\ref{12}) related to BT of equation (2), we would like to
present two sets of negative pKdV hierarchies and their
corresponding bilinear forms are also constructed only by the
transformation (\ref{7}). \\
\emph{Case 1.} The first kind of negative pKdV
hierarchy can be obtained, reading
\begin{eqnarray}
u_{t_{-N}}=-\sum^{N}_{i=1} \exp(\int{u-u_i}\rmd x), \label{16}\\
u_x+u_{i,x}=-2\lambda_i+\frac{(u-u_i)^2}{2},\quad i=1,2,...,N,
\label{17}
\end{eqnarray}
where $\lambda_i$ is arbitrary constant.

In particular, when  $N=1$, one has the first equation of negative
pKdV hierarchy, namely
\begin{eqnarray}
2u_{xxt}u_t-4u_xu^2_t-u^2_{xt}-4\lambda_1u^2_t=0.\label{18}
\end{eqnarray}
Here we have instead $t_{-1}$ with $t$ for simplicity. It is well
known that the first negative flow in the KdV hierarchy is linked to
the Camassa-Holm equation via a hodograph transformation \cite{24}
or can be reduced to the sinh-Gordon/sine-Gordon/Liouville equations
\cite{25}. Here we transform equation (\ref{18}) into sine-Gordon
and Liouville equations.

In fact, by setting $\beta\equiv \beta(x,t)=-u_t$, we can rewrite
equation (\ref{18}) in the form
\begin{eqnarray}
\beta_x=\left(-\frac{\beta_{xx}}{2\beta}+\frac{\beta^2_x}{4\beta^2}\right)_t,
\label{19}
\end{eqnarray}
which can be integrated once  with respect to $x$  to give
\begin{eqnarray}
\beta(\ln{\beta})_{xt}+\beta^2=\beta_0(t), \label{20}
\end{eqnarray}
where $\beta_0(t)$ is an arbitrary function of $t$.

As it is reported in Ref.\cite{24}, for non-zero $\beta_0(t)$, one
can rescale $\beta$ to $\sqrt{\beta_0(t)}\beta$, redefine $t$ as
$t/\sqrt{\beta_0(t)}$ and set $\beta=\exp ({i\eta})$ to give the
sine-Gordon equation
\begin{eqnarray}
\eta_{xt}=\sin{\eta},\label{21}
\end{eqnarray}
while for $\beta_0(t)=0$, by setting $\beta=-\exp({\eta})$, equation
(\ref{20}) becomes the Liouville equation
\begin{eqnarray}
\eta_{xt}=\rm{e}^{\eta}. \label{22}
\end{eqnarray}
\textbf{\emph{Remark 1.}} The quantity
$-\frac{\beta_{xx}}{2\beta}+\frac{\beta^2_x}{4\beta^2}\equiv A$ in
the right hand side of equation (\ref{19}) can be given in terms of
a Miura transformation
\begin{eqnarray}
A=-\theta_x-\theta^2, \quad \theta=\frac{\beta_x}{2\beta}.
\label{23}
\end{eqnarray}
Furthermore, by virtue of the dependent variable transformation
\begin{eqnarray}
u=-2\frac{\psi_x}{\psi}, \qquad u_i=-2\frac{\psi_{i,x}}{\psi_i},
\qquad (i=1,2,...,N)\label{24}
\end{eqnarray}
the negative pKdV hierarchy (\ref{16})-(\ref{17}) is directly
transformed into its bilinear form
\begin{eqnarray}
D_x D_{t_{-N}}\psi \cdot \psi =\sum^N_{i=1}\psi^2_i, \label{25}\\
(D^2_x-\lambda_i) \psi\cdot \psi_i= 0, \quad i=1,2,...,N. \label{26}
\end{eqnarray}
\emph{Case 2.} For the nonlocal symmetry (\ref{12}) being dependent
with parameter  $\lambda$, we may derive the second kind of negative
pKdV hierarchy by expanding the dependent variable in power series of
$\lambda$. In this case, we have
\begin{eqnarray}
\left. u_{t_{-N}}=-\frac{1}{N!} \left(\frac{\partial^{N}
\exp(\int{u-u_1}\rmd x)}{\partial \lambda^{N}} \right)
\right|_{\lambda=0},\label{27}\\
u_x+u_{1,x}=-2\lambda+\frac{(u-u_1)^2}{2}. \label{28}
\end{eqnarray}
Under the transformation $u=-\frac{2\psi_x}{\psi}$ and
$u_1=-\frac{2\psi_{1,x}}{\psi_1}$, the negative pKdV hierarchy
(\ref{27})-(\ref{28}) becomes
\begin{eqnarray}
\left.D_x D_{t_{-N}} \psi \cdot \psi=\frac{1}{N!}
\left(\frac{\partial ^{N}\psi^2_1}{\partial \lambda ^{N}}\right)
\right|_{\lambda=0},\label{29}
\\ (D^2_x-\lambda)\psi\cdot \psi_1=0. \label{30}
\end{eqnarray}
Let $\psi_1=\psi_1(\lambda)$ have a formal series form
\begin{eqnarray}
\psi_1=\sum^{\infty}_{i=0} \bar{\psi}_{i}\lambda ^{i}, \label{31}
\end{eqnarray}
where $\bar{\psi}_{i}$ is $\lambda$ independent. Then
(\ref{29})-(\ref{30}) can be rewritten as
\begin{eqnarray}
D_x D_{t_{-N}} \psi \cdot
\psi=\sum^{N}_{k=0}\bar{\psi}_k\bar{\psi}_{N-k},\label{32}\\
D^2_x \psi \cdot \bar{\psi}_k=\psi \bar{\psi}_{k-1} \qquad
(k=0,1,...,N) \label{33}
\end{eqnarray}
with $\bar{\psi}_{-1}=0.$

The negative pKdV hierarchy in bilinear form (\ref{32})--(\ref{33})
is just the special situation of the bilinear negative KP hierarchy
for
$\partial_y=0$ in Ref.\cite{22}. From this observation, we have the following remark: \\
{\bf {\em {Remark 2.}}} The second negative pKdV hierarchy shown by
(\ref{27})-(\ref{28})
 is a potential form of a known negative KdV hierarchy given by other methods, say,
 the inverse recursion operator \cite{4}, Lax operator \cite{23}, and the Guthrie's approach
 \cite{Gu}.

\section{Localization of the nonlocal symmetries}

We know that the  Lie point symmetries \cite{2,3} can be applied to
construct finite symmetry transformation and group invariant
solutions for DEs, whereas the calculations are invalid for the
nonlocal symmetries. So it is anticipant to turn the nonlocal
symmetries into local ones, especially into  Lie point symmetries.
In order to make the nonlocal symmetry localized, one may extend the
original system to a closed prolonged system by introducing some
additional dependable variables \cite{12,13,14} to eliminate
integration and differentiation.

Fortunately,  starting from  the nonlocal symmetry (\ref{12}), the
prolongation is found to be closed when another two dependent
variables $v\equiv v(x,t)$ and $g\equiv g(x,t)$ are introduced by

\begin{eqnarray}
\eqalign{v_x=u-u_1, \qquad v_t=2(u-u_1)(u_x-2\lambda)-2u_{xx}, \cr
g_x=\rme^v, \qquad \qquad g_t=-e^v[2u_x+8\lambda-(u-u_1)^2].}
\label{43}
\end{eqnarray}

Now the prolonged equations (2), (3), (4) and (\ref{43}) contain
four dependable variables $u$, $u_1$, $v$ and $g$, whose
corresponding symmetries are
\begin{eqnarray}
\sigma_u=\rme^{v},\quad \sigma_{u_1}=0,\quad \sigma_v=g,\quad
\sigma_g=\frac{1}{2}g^2. \label{44}
\end{eqnarray}
\textbf{\emph{Remark 3.}} What is more interesting here is that the symmetry $\sigma^g$ shown
in (\ref{44}) implies the
auxiliary dependent variable $g$ satisfies
\begin{equation}
g_t=\{g; x\}g_x+6\lambda g_x, \ \{g; x\}\equiv \frac{g_{xxx}}{g_x}-\frac32\frac{g_{xx}^2}{g_x^2},
\label{skdv}
\end{equation}
which is just the Schwartz form of the KdV (SKdV) equation (1). This
may provide us with a new way to seek for the Schwartz forms of DEs,
especially for the discrete integrable models, without using
Painlev\'e analysis.

Due to (\ref{44}), the symmetry vector of the prolonged system has
the form
\begin{eqnarray}
V=\rme^{v} \frac{\partial}{\partial u}+0 \frac{\partial}{\partial
u_1}+g\frac{\partial} {\partial v} +\frac{1}{2}g^2
\frac{\partial}{\partial g}.
\end{eqnarray}
Then, by solving the following initial value problem
\begin{eqnarray}
\eqalign{\frac{d\bar{u}}{d
\epsilon}=\rme^{\bar{v}},\quad\frac{d\bar{u}_1}{d \epsilon}=0,\quad
\frac{d\bar{v}}{d \epsilon}=\bar{g},\quad \frac{d\bar{g}}{d
\epsilon}=\frac{1}{2}\bar{g}^2, \cr \bar{u}|_{\epsilon=0}=u, \quad
\bar{u}_1 |_{\epsilon=0}=u_1, \quad\bar{v}|_{\epsilon=0}=v,
\quad\bar{g}|_{\epsilon=0}=g, }
\end{eqnarray}
the finite transformation can be written out as follows
\begin{eqnarray}
\bar{u}=u+\frac{2\epsilon}{2-\epsilon g}\rme^v, \quad \bar{u}_1=u_1,
\quad \bar{v}=v+2\ln\frac2{2-\epsilon g}, \quad
\bar{g}=\frac{2}{2-\epsilon g}g.\label{47}
\end{eqnarray}
\textbf{\emph{Remark 4.}} The original BT (3) and (4) in itself
suggests a finite transformation from one solution $u$ to another
one $u_1$ and then the new BT (\ref{47}) obtained via (\ref{12})
will arrive at a third solution $\bar u$. Actually, the finite
transformation (\ref{47}) is just the so-called Levi transformation
\cite{26}. The result of this paper shows the fact that two kinds of
BT possess the same infinitesimal form (\ref{12}).

Now by force of the finite  transformation (\ref{47}), one can get
new solution from any initial solution. For example, it is easy to
solve an initial solution of prolonged equation system (2), (3), (4)
and (\ref{43}), namely
\begin{eqnarray}
\eqalign{u=c,\qquad u_1=c+2\sqrt{\lambda}\tanh{\zeta}, \qquad
v=-\ln(\tanh^2{\zeta}-1), \cr
g=\frac{\sinh(2\zeta)}{4\sqrt{\lambda}} -\frac{x}{2}+6\lambda t+c_0,
\qquad \zeta=\sqrt{\lambda}(-x+4\lambda t),} \label{48}
\end{eqnarray}
Where $\lambda$, $c$ and $c_0$ are three arbitrary constants.

Starting from this  original solution (\ref{48}), a new solution of
equation (2) can be presented immediately from (\ref{47}):
\begin{eqnarray}
 \bar{u}=c-\frac{8\sqrt{\lambda}\epsilon
\cosh^2{\zeta}}{8\sqrt
{\lambda}-\epsilon[\sinh(2\zeta)-2\sqrt{\lambda}( x-12\lambda
t-2c_0)]},\label{49}
\end{eqnarray}
which then gives the corresponding solution of KdV equation
\begin{eqnarray}
\fl \bar{\omega}=\bar{u}_x=16\lambda \epsilon \cdot
\frac{[\cosh(2\zeta)+1]\epsilon+\sqrt{\lambda}\sinh(2\zeta)
[4+\epsilon( x-12\lambda t-2c_0)] }{[8\sqrt
{\lambda}-\epsilon(\sinh(2\zeta)-2\sqrt{\lambda}( x-12\lambda
t-2c_0))]^2}
\end{eqnarray}
with $\zeta=\sqrt{\lambda}(-x+4\lambda t)$.

Besides obtaining new solutions from old ones, symmetries can be
applied to get special solutions that  are  invariant under  the
symmetry transformations by reducing dimensions of a partial
differential equation. To find more similarity reductions of
equation (2), we will study  Lie point symmetries of the whole
prolonged equation system instead of the single equation (2).
Suppose equations (2), (3), (4) and (\ref{43}) be invariant under
the infinitesimal transformations
\begin{eqnarray}
u\rightarrow u+\epsilon \sigma, \qquad u_1\rightarrow u_1+\epsilon
\sigma_1, \qquad v\rightarrow v+\epsilon \sigma_2,\qquad
g\rightarrow g+\epsilon \sigma_3,\nonumber
\end{eqnarray}
with
\begin{eqnarray}
\eqalign{\sigma=X(x,t,u,u_1,v,g)u_x+T(x,t,u,u_1,v,g)u_t-U(x,t,u,u_1,v,g),
\cr
\sigma_1=X(x,t,u,u_1,v,g)u_{1,x}+T(x,t,u,u_1,v,g)u_{1,t}-U_1(x,t,u,u_1,v,g),\cr
\sigma_2=X(x,t,u,u_1,v,g)v_x+T(x,t,u,u_1,v,g)v_t-V(x,t,u,u_1,v,g),\cr
\sigma_3=X(x,t,u,u_1,v,g)g_x+T(x,t,u,u_1,v,g)g_t-G(x,t,u,u_1,v,g).}\label{51}
\end{eqnarray}
Then substituting the expressions (\ref{51}) into the symmetry
equations of equations (2), (3), (4) and (\ref{43})
\begin{eqnarray}
\eqalign{\sigma_t+\sigma_{xxx}-6u_x\sigma_x=0, \cr
\sigma_{1,x}+\sigma_x-(\sigma-\sigma_1)(u-u_1)=0, \cr
\sigma_{1t}-\sigma_{xxx}+2(u-u_1)\sigma_{xx}+2(\sigma-\sigma_1)u_{xx}-
[4\lambda+(u-u_1)^2-2u_x]\sigma_x \\
+2(\sigma-\sigma_1)(u-u_1)(2\lambda-u_x)=0,\cr
\sigma_{2,x}-\sigma+\sigma_1=0,\cr
\sigma_{2t}+2\sigma_{xx}+2(u_1-u)\sigma_x+2(\sigma_1-\sigma)(u_x-2\lambda)=0,\cr
\sigma_{3,x}-\rme^v\sigma_2=0,\cr
\sigma_{3t}+2\rme^v[\sigma_x+(u_1-u)(\sigma-\sigma_1)-\frac{1}{2}(u-u_1)^2\sigma_2
+(4\lambda+u_x)\sigma_2]=0,} \label{52}
\end{eqnarray}
and collecting together the coefficients of partial derivatives of
dependent variables, it yields a system of overdetermined linear
equations for the infinitesimals $X$, $T$, $U$, $U_1$, $V$ and $G$,
which can be solved by virtue of \emph{Maple} to give
\begin{eqnarray}
\eqalign{X(x,t,u,u_1,v,g)=c_1(x+12\lambda t)+c_5,\cr
T(x,t,u,u_1,v,g)=3c_1t+c_2,\cr U(x,t,u,u_1,v,g)=-c_1(2\lambda
x+u)+2c_4 \rme^v+c_3,\cr U_1(x,t,u,u_1,v,g)=-c_1(2\lambda
x+u_1)+c_3, \cr V(x,t,u,u_1,v,g)=-c_1+2c_4g+c_6,\cr
G(x,t,u,u_1,v,g)=c_4g^2+c_6g+c_7,} \label{53}
\end{eqnarray}
where $c_i (i=1...7)$ are seven arbitrary constants. When
$c_1=c_2=c_3=c_5=c_6=c_7=0$, the reduced symmetry is just
(\ref{44}).

To give the group invariant solutions, we would like to solve
symmetry constraint conditions $\sigma=0$ and
$\sigma_{i}=0(i=1,2,3)$ defined by (\ref{51}) with (\ref{53}), which
is equivalent to solve the following characteristic equation
\begin{eqnarray}
\eqalign{\frac{\rmd x}{c_1(x+12\lambda t)+c_5}=\frac{\rmd
t}{3c_1t+c_2}=\frac{\rmd u}{-c_1(2\lambda x+u)+2c_4 \rme^v+c_3} \cr
=\frac{\rmd u_1}{-c_1(2\lambda x+u_1)+c_3}=\frac{\rmd
 v}{-c_1+2c_4g+c_6}=\frac{\rmd g}{c_4g^2+c_6g+c_7}.}\label{54}
\end{eqnarray}
Two nontrivial similar reductions under consideration $c_4\neq0$ are
presented and  substantial group invariant solutions are found in
the follows.\\
\emph{Case 1:} $c_1 \neq 0$ and $c^2_6-4c_4c_7\neq0$.

Without loss of generality, we let $c_1\equiv 1$. For simplicity, we
introduce arbitrary constants $a_4$ and $a_7$ to replace $c_4$ and
$c_7$ by $a^2_4=c^2_6-4c_4c_7$ and $a_7=-a^2_4/(16c_4)$, then after
solving equation (\ref{54}), we have
\begin{eqnarray}
\eqalign {u=-\lambda x+3\lambda^2 t+c_3+c_5\lambda-3c_2\lambda^2+
(3t+c_2)^{-\frac{1}{3}}[U(\xi)\\ \qquad
-\frac{a_4}{4a_7}\exp(V(\xi)-G(\xi)) \tanh B ],\cr u_1=-\lambda
x+3\lambda^2
t+c_3+c_5\lambda-3c_2\lambda^2+\frac{U_1(\xi)}{(3t+c_2)^{\frac{1}{3}}},\cr
v=-\frac{1}{3}\ln(3t+c_2)-G(\xi)+V(\xi)-2\ln{\cosh B},\cr
g=\frac{8a_7}{a_4}[\tanh B+\frac{c_6}{a_4}]}\label{55}
\end{eqnarray}
with  $B= a_4(3G(\xi)+\ln(3c_1t+c_2))/6$ and $\xi=(x-6\lambda
t+c_5-6c_2\lambda)/(3t+c_2)^{\frac{1}{3}}.$

Here, $U(\xi)$, $U_1(\xi)$, $V(\xi)$, $G(\xi)$ and $\xi$ represent
five group invariants and substituting (\ref{55}) into the prolonged
equations system  gives the following reduced  equations
\begin{eqnarray}
H_{\xi\xi}=\frac{1}{2}\frac{H^2_{\xi}}{H}+4a_7H^2-\xi
H-\frac{a^2_4}{32a^2_7H},\label{56}
\end{eqnarray}
\begin{eqnarray}
\eqalign{U_1(\xi)=\frac{a_7H^2_{\xi}}{H}-4a^2_7H^2+2a_7\xi
H-\frac{\xi^2}{4}-\frac{a^2_4}{16a_7H}\cr
U(\xi)=U_1(\xi)-\frac{H_{\xi}}{H},\quad V(\xi)=G(\xi)-\ln(H),\quad
G_{\xi}(\xi) =\frac{1}{4a_7H}} \label{57}
\end{eqnarray}
with $H\equiv H(\xi).$ One can see that whence $H$ is solved from
equation (\ref{56}), two new group invariant solutions $u$ and $u_1$
of equation (2) would be immediately obtained through equations
(\ref{55}) and (\ref{57}).

Moreover, by making a further transformation \cite{27}
\begin{eqnarray}
H(\xi)=\frac{1}{2a_7}(P_{\xi}+P^2+\frac{\xi}{2}), \qquad P\equiv
P(\xi),
\end{eqnarray} equation (\ref{56})  can be converted  into the
second  Painlev\'{e} equation ${\rm{P}_{II}}$, reading
\begin{eqnarray}
P_{\xi\xi}=2P^3+\xi P+\alpha, \label{59}
\end{eqnarray}
with $\alpha=-(a_4+1)/2$. Now, every known solution of
${\rm{P}_{II}}$ (\ref{59}) will generate two new group invariant
solutions of equation (2), and then two new solutions of KdV
equation (1) denoted as $\omega_1$ and $\omega_2$  can be given
directly after one derivative with respect to $x$ for $u_1$ and $u$
\begin{eqnarray}
\omega_1=\frac{1}{(3t+c_2)^{\frac{2}{3}}}(P_{\xi}+P^2)-\lambda, \label{60}\\
\fl \omega_2=\frac{1}{(3t+c_2)^{\frac{2}{3}}}[-\frac{a^2_4}{2F^2}
{\rm{sech}} ^2{R_1}+(\frac{2a_4P}{F}- \frac{a^2_4}{F^2})\tanh
{R_1}+\frac{2a_4P}{F}+P_{\xi}-P^2]+\lambda, \label{61}
\end{eqnarray}
where
\begin{eqnarray}
\fl F\equiv F(\xi)=2P_{\xi}+2P^2+\xi, \quad
R_1=\frac{1}{6}a_4[\ln(3t+c_2)+3G(\xi)],\quad
G_{\xi}(\xi)=\frac{1}{2P_{\xi}+2P^2+\xi}, \nonumber
\end{eqnarray}
and $P$ satisfies ${\rm{P}_{II}}$ (\ref{59}) with
$\alpha=-(a_4+1)/2$.

It is known that the generic solutions of ${\rm{P}_{II}}$ are
meromorphic functions and more information about ${\rm{P}_{II}}$ is
provided in Ref.\cite{28}, saying: (1) For every $\alpha=N \in Z$,
there exists a unique rational solution of ${\rm{P}_{II}}$; (2) For
every $\alpha=N+\frac{1}{2}$, with $N \in Z$, there exists a unique
one-parameter family of classical solutions which are expressible in
terms of Airy functions; (3) For all other values of $\alpha$, the
solution of ${\rm{P}_{II}}$ is transcendental.

For example, when $\alpha=1$ $(a_4=-3)$, ${\rm{P}_{II}}$ (\ref{59})
possesses a simple rational solution $P(\xi)=-{1}/{\xi}$, which
leads the solutions (\ref{60}) and (\ref{61}) to
\begin{eqnarray}
\tilde{\omega}_1=\frac{2}{(x-6\lambda t+c_5-6c_2\lambda)^2
}-\lambda, \label{62}
\end{eqnarray}
and
\begin{eqnarray}
\fl \tilde{\omega}_2 =-[x^6-36tx^5+(540t^2-6)x^4-(4320t^3-168t-2)x^3+36t(540t^3-48t-1)x^2 \nonumber\\
\qquad \fl -(46656t^5 -7776t^3-216t^2-144t-12)x+46656t^6-12960t^4-432t^3 \nonumber\\
\qquad \fl -720t^2-48t+1] /[x^3-18x^2t+108xt^2-(6t+1)(36t^2-6t-1)]^2
\label{63}
\end{eqnarray}
In the formulation (\ref{63}), we have made $c_2=0$, $c_5=0$ and
$\lambda=1$ because the original expression is much too complicated.
The simple rational solutions of PII will yield   abundant rational
solutions of KdV equation.

When $\alpha=\frac{1}{2}$  $(a_4=-2)$, ${\rm{P}_{II}}$ (\ref{59})
has a solution expressed by $Airy$ function
\begin{eqnarray}
P(\xi)=2^{-\frac{1}{3}}\frac{3{\rm{Ai}}(1,-2^{-\frac{1}{3}}\xi)-\sqrt{3}{\rm{Bi}}(1,
-2^{-\frac{1}{3}}\xi)}
{3{\rm{Ai}}(-2^{-\frac{1}{3}}\xi)-\sqrt{3}{\rm{Bi}}(-2^{-\frac{1}{3}}\xi)}.
\label{64}
\end{eqnarray}
For simplicity, we convert equation (\ref{64}) into the equivalent
form
\begin{eqnarray}
P(\xi)=\frac{\sqrt{2}\xi^{\frac{3}{2}}{\rm{J}}({\frac{4}{3}},\frac{\sqrt{2}}{3}\xi^{\frac{3}{2}})
-2{\rm{J}}({\frac{1}{3}},\frac{\sqrt{2}}{3}\xi^{\frac{3}{2}})} {\xi
{\rm{J}}({\frac{1}{3}},\frac{\sqrt{2}}{3}\xi^{\frac{3}{2}})},
\label{65}
\end{eqnarray}
where ${\rm{J}}(n,\xi)$ is the first kind of Bessel function.
Substituting (\ref{65}) into (\ref{60}) and (\ref{61}) with
$c_2=c_5=0 $ and $\lambda=1$ (or else the formulae are too long to
written down here), two exact solutions of KdV equation are obtained
as follows:
\begin{eqnarray}
\omega_1'=\frac{x-12t}{6t}+\frac{x-6t}{3t}\frac{{\rm{J}}^2_2}{{\rm{J}}^2_1},
\label{66}\\
\omega_2'=-\Psi/\Omega \label{67}
\end{eqnarray}
with

\begin{eqnarray} \fl
\Psi=32\sqrt{t}(x-6t)^4[(x(x-6t)^2-12t){\rm{J}}^6_1+x(x-6t)^2{\rm{J}}^6_2]
+128\sqrt{6}t(x-6t)^{\frac{11}{2}}\cdot({\rm{J}}^5_1{\rm{J}}_2\nonumber\\
\fl \qquad +{\rm{J}}^5_2{\rm{J}}_1)+96\sqrt{t}(x-6t)^4
[(x(x-6t)^2-4t){\rm{J}}^4_1{\rm{J}}^2_2+x(x-6t)^2{\rm{J}}^4_2{\rm{J}}^2_1]-72\cdot
2^{\frac{1}{3}}\sqrt{t}\nonumber\\ \fl \qquad \cdot (x-6t)^2
\cdot[(x(x-6t)^2-4t)\cdot
{\rm{J}}^4_1+x(x-6t)^2{\rm{J}}^4_2]+256\sqrt{6}t
(x-6t)^{\frac{11}{2}}{\rm{J}}^3_1{\rm{J}}^3_2\nonumber\\ \fl \qquad
-192\sqrt{3}\cdot 2^{\frac{5}{6}}t(x-6t)^{\frac{7}{2}} \cdot
({\rm{J}}^3_1{\rm{J}}_2+{\rm{J}}^3_2{\rm{J}}_1)+54\cdot
2^{\frac{2}{3}} \cdot
x\sqrt{t}(x-6t)^2({\rm{J}}^2_1+{\rm{J}}^2_2)\nonumber\\ \fl \qquad
-144\cdot
2^{\frac{1}{3}}x\sqrt{t}(x-6t)^4{\rm{J}}^2_1{\rm{J}}^2_2+144\sqrt{3}
\cdot2^{\frac{1}{6}}t(x-6t)^{\frac{3}{2}}
{\rm{J}}_1{\rm{J}}_2-27x\sqrt{t},\nonumber
\end{eqnarray}
and
\begin{eqnarray}
\fl
\Omega=6t^{\frac{3}{2}}[2^{\frac{5}{3}}(x-6t)^2({\rm{J}}^2_1+{\rm{J}}^2_2)-3]
[8\cdot2^{\frac{1}{3}}(x-6t)^4
({\rm{J}}^2_1-{\rm{J}}^2_2)^2-12\cdot2^{\frac{2}{3}}(x-6t)^2
({\rm{J}}^2_1+{\rm{J}}^2_2)+9],\nonumber
\end{eqnarray}
where we denote
${\rm{J}}_1={\rm{J}}(\frac{1}{3},\frac{\sqrt{6}}{9}\frac{(x-6t)^{\frac{3}{2}}}{\sqrt{t}}),\quad
{\rm{J}}_2={\rm{J}}(-\frac{2}{3},\frac{\sqrt{6}}{9}\frac{(x-6t)^{\frac{3}{2}}}{\sqrt{t}})
$.

Then continue to do the same, sequences of rational solutions and
Bessel (Airy) function solutions for KdV equation will be easily
constructed. Furthermore, by selecting suitable parameters in this
kind of similar reduction, we may discover more unknown  exact
solutions among interaction solitons and Painlev\'{e} waves of KdV equation.
\\
\emph{Case 2:} $c_1=0$ and $c_2\neq 0$.

Firstly, it is convenient to replace $c_4$ and $c_5$ with $a_4$ and
$k$ by $a^2_4=c^2_6-4c_4c_7$ and $k={c_5}/{c_2}$, and it follows the
results from equation (\ref{54}), saying
\begin{eqnarray}
\eqalign{
u=\frac{c_3}{c_2}t+U(z)+\frac{c_3}{c_2}G(z)-\frac{(a^2_4-c^2_6)
}{a_4c_7}\rm{e}^{V(z)} \tanh[\frac{a_4(t+G(z))}{2c_2}],\cr
u_1=\frac{c_3}{c_2}t+U_1(z),\cr
g=\frac{2c_7}{a^2_4-c^2_6}[c_6+a_4\tanh(\frac{a_4(t+G(z))}{2c_2})],
\cr v=V(z)-2\ln{\cosh[\frac{a_4(t+G(z))}{2c_2}]}} \label{68}
\end{eqnarray}
with $z=x-kt$. Substituting (\ref{68}) into equations (2), (3), (4)
and (\ref{43}) and redefining the parameters for the sake of
simplicity, we notice that the new group invariants $U(z)$,
$U_1(z)$, $V(z)$ and $G(z)$ are subject to
\begin{eqnarray}
W^2_z-a^2_2W^4-a_3W^3+a_5W^2- a_7W=0, \label{69}
\end{eqnarray}
\begin{eqnarray}
\eqalign{ U_{1z}(z)=\frac{a_7}{2W}-\lambda-\frac{a_5}{4},\cr
U(z)=U_1(z)+\frac{W_z}{W}+(3\lambda^2+\frac{\lambda
a_5}{2}-\frac{a^2_5}{16}+\frac{a_3a_7}{4})G,\cr V(z)=\ln(W), \cr
G_z(z)=\frac{W}{a_7}} \label{70}
\end{eqnarray}
with $W\equiv W(z)$ and
 $a_2=\frac{a^2_4-c^2_6}{a_4c_7}$,
$a_3=\frac{(a^2_4-c^2_6)(c_2k^2+48c_2\lambda^2-16c_2k\lambda-4c_3)}{a^2_4c_7}$,
$a_5=2k-12\lambda$, $a_7=\frac{a^2_4c_7}{c_2(a^2_4-c^2_6)}$.\\
{\bf \em {Remark 5}}. The case $c_1=0$  here is interesting. From
equation (\ref{69}), we know that $W$ can be expressed as an
elliptic integration and can be expressed by means of Jaccobi
elliptic functions. Whence $W$ is fixed from (\ref{69}), all the
other quantities are given simply given by differentiation or
integration. The first equation of (\ref{68}) implies the important
byproduct, the explicit exact interaction between cnoidal periodic
wave and kink soliton.

A simple example of this case can be obtained by using the simplest Jacobi Elliptic function expansion method which leads to
\begin{eqnarray}
W(z)=\frac{a_3}{4a^2_2}[{\rm{sn}}(\frac{a_3z}{4a_2n},n)-1] \label
{71}
\end{eqnarray}
with the constraint conditions
$a_5=\frac{a^2_3(1-5n^2)}{16n^2a^2_2}$ and
$a_7=\frac{a^2_3(n^2-1)}{32n^2a^4_2}$ in equation (\ref{69}), where
$n$ is the modulus of the Jacobian elliptic function $\rm{sn}$.

After solving equation (\ref{70}) with the given solution (\ref{71})
and taking the results into (\ref{68}), two exact solutions of the KdV
equation are obtained
\begin{eqnarray}
\omega_3=\frac{a^2_3(1-n^2)}{16a^2_2n^2(Y+1)}+
\frac{a^2_3(5n^2-1)}{64a^2_2n^2}-\lambda, \label {72}\\
\fl \omega_4=\frac{a^2_3(Y+1)^2}{32a^2_2}\tanh^2{R_2}
+\frac{a^2_3\sqrt{Y^2-1}\sqrt{n^2Y^2-1}} {16na^2_2}\tanh{R_2}+
\frac{2a^2_3(Y^2-2Y)}{a^2_2}\nonumber
\\-\frac{a^2_3(n^2+1)}{n^2a^2_2}-64\lambda, \label{73}
\end{eqnarray}
where we have
$$R_2=\frac{a^3_3(n^2-1)(t+a_6)}{64a^3_2n^2}+\frac{1}{4}\ln\left({\frac{2n}{2n^2Y^2+2n
\sqrt{Y^2-1}\sqrt{n^2Y^2-1}-n^2-1}}\right)$$$$+\frac{1}{2}n
\int^{Y}_0{\frac{1}{\sqrt{1-t^2}\sqrt{1-n^2t^2}}}dt, \quad
Y=\rm{sn}\left(\frac{a_3(192\lambda
a^2_2\emph{n}^2-5a^2_3\emph{n}^2+a^2_3)\emph{t}}{128\emph{n}^3a^3_2}-\frac{a_3\emph{x}}
{4a_2\emph{n}},\emph{n}\right),\quad
$$ and $a_2$, $a_3$, $a_6$ and $\lambda$ are four arbitrary
constants.

\section{Integrable models from nonlocal symmetry with B\"acklund transfor-
mation}

To find new integrable models is another important application of
the symmetry study. Symmetry constraint method is one of the most
powerful tools to give out new integrable models from known ones.
Especially, casting symmetry constraint condition to  Lax pair of
soliton equations, one can obtain many other integrable models. In
this section, we would like to combine the nonlocal symmetry with BT
of pKdV equation to give some  integrable models both in lower and
higher dimensions.

Let every pair $(u, u_i)$ ($i=1,2,...,N$) satisfy the following BT
\begin{eqnarray}
u_x+u_{i,x}=-2\lambda_i+\frac{(u-u_i)^2}{2},
\label {74}\\
u_t+u_{i,t}=2u^2_{x}+2u^2_{i,x}+2u_xu_{i,x}-(u-u_i)(u_{xx}-u_{i,xx}),\label{75}
\end{eqnarray}
and the corresponding nonlocal symmetry of $u$ reads
$\sigma^{i}=\exp(\int{u-u_i}\rmd x)$ for $i=1,2,...,N$.

\subsection{Finite-dimensional  integrable systems}

In general, every one symmetry of a higher dimensional model can
lead the original one to its lower form. Now, considering
\begin{eqnarray}
u_x=\sum_{i=1}^N a_i\exp(\int{u-u_i}\rmd x) \label {76}
\end{eqnarray}
as a generalized symmetry constraint condition and acting it on the
$x$-part of the BT (3), we firstly give the finite dimensional
$(N+1)$-component integro-differential system
\begin{eqnarray}
\eqalign{u_x=\sum_{i=1}^N {a_i\exp(\int{u-u_i}\rmd x)},\cr
u_x+u_{i,x}=-2\lambda_i+\frac{(u-u_i)^2}{2},\quad i=1,2,...,N,}
\label{77}
\end{eqnarray}
where every $a_i$ and $\lambda_i$ are arbitrary constants. For
further simplification,  making  $u_i=u-(\ln{w_{ix}})_x$, then the
constraint condition (\ref{76}) becomes
\begin{eqnarray}
u=\sum_{m=1}^N a_m w_m, \label{78}
\end{eqnarray}
which  transforms  (\ref{77}) into the  N-component differential
system
\begin{eqnarray}
\fl 2w_{ixxx}w_{ix}-4(\sum^{N}_{m=1}a_m
w_{mx})w^2_{ix}-w^2_{ixx}-4\lambda_iw^2_{ix}=0, \qquad i=1,2,...,N.
\label{79}
\end{eqnarray}
Taking $s_i=w_{ix}$, we rewrite equation (\ref{79}) as
\begin{eqnarray}
2s_{ixx}s_i-4(\sum^{N}_{m=1}a_ms_m)s^2_{i}-s^2_{ix}-4\lambda_is^2_{i}=0,
 \quad i=1,2,...,N.\label{80}
\end{eqnarray}
Making $s_i=b_i q^2_i$, equation (\ref{80}) is equivalent to the
downward integrable system
\begin{eqnarray}
q_{ixx}-(\sum^{N}_{m=1}c_m q^2_m)q_i-\lambda_i q_i=0, \quad
i=1,2,...,N\label{81}
\end{eqnarray}
with $c_i=a_ib_i$ being arbitrary constant.

 On the other hand,  by the same symmetry constraint  (\ref{76}) and
the $t$-part of the BT (4), we can construct another set of
integrable system
\begin{eqnarray}
\eqalign{u_x=\sum_{i=1}^N {a_i\exp(\int{u-u_i}\rmd x)},\cr
 u_t+u_{i,t}=2u^2_{x}+2u^2_{i,x}+2u_xu_{i,x}-(u-u_i)(u_{xx}-u_{i,xx})
,\quad i=1,2,...,N.} \label{82}
\end{eqnarray}
Considering the similarity transformations of dependent variables
done in $x$-part, after a series of tedious substitutions, equation
(\ref{82}) becomes
\begin{eqnarray}
\fl q_iq_{ixt}-q_{it}q_{ix}+(2\sum_{m=1}^N c_m
q^2_m-4\lambda_i)q^2_{ix}-4q_iq_{ix}\sum_{m=1}^N
c_mq_mq_{mx}-2q^2_i(\sum_{m=1}^N c_mq^2_m)^2 \nonumber\\
\fl  +2\lambda_i q^2_i \sum_{m=1}^N c_m q^2_m
+4\lambda^2_iq^2_i+2q^2_i\sum_{m=1}^N c_m(q^2_{mx}+q_mq_{mxx})=0,
\qquad i=1,2,...,N. \label{83}
\end{eqnarray}
Taking equation (\ref{81}) into account, equation (\ref{83}) can be
integrated once about $x$ to give $N$-component integrable system
\begin{eqnarray}
q_{it}=-2\sum_{m=1}^N c_m q_mq_{mx}q_i+2\sum_{m=1}^N c_mq^2_m
q_{ix}-4\lambda_iq_{ix}, \qquad i=1,2,...,N. \label{84}
\end{eqnarray}

In fact, equations (\ref{81}) and (\ref{84}) are essentially  the
canonical equation $(F_0)$ and $(F_1)$ respectively [16], saying
\begin{eqnarray}
(F_0): \qquad q_{ix}=p_i, \qquad p_{ix}= (\sum_{m=1}^N c_m
q^2_m)q_i+\lambda_i q_i. \label{F0}
\end{eqnarray}
\begin{equation}
\label{F1} (F_1): \left\{
\begin{array}{rl}
q_{it}&=-2 (\sum_{m=1}^N c_m p_m q_m) q_i+2(\sum_{m=1}^N c_m q^2_m)
p_i-4\lambda_ip_i,\\
p_{it}&=2(\sum_{m=1}^N c_m p_m q_m) p_i-2(\sum_{m=1}^N c_m
p^2_m)q_i-4\lambda^2_iq_i\\
&-2\lambda_i(\sum_{m=1}^N c_m q^2_m)q_i-2(\sum_{m=1}^N \lambda_m c_m
q^2_m)q_i.
\end{array}
\right.
\end{equation}

It should be stressed here that the finite integrable systems
(\ref{81}) and (\ref{84}) reobtained via this way are just the
remarkable results given by Cao in Ref.\cite{16} through the
nonlinearization method, both of which have been proved completely
integrable in Liouville sense. Thanks to these  finite integrable
systems (\ref{F0}) and (\ref{F1}), the original high dimensional KdV
equation  would be solved.

\subsection{Infinite-dimensional  integrable systems}

For getting some higher dimensional integrable models, one may
introduce some internal parameters \cite{6,8,20}. Here, we would
like to use the internal parameter dependent symmetry constraints on
BT to construct two sets of infinite-dimensional integrable systems.

It is obvious that equation (2) is invariant under the internal
parameter translation, say $y$ translation,  so we can view
\begin{eqnarray}
u_y=\sum_{i=1}^N a_i\exp(\int{u-u_i}\rmd x)\label{85}
\end{eqnarray}
as a new symmetry constraint condition.

Firstly, imposing (\ref{85}) on the $x$-part of the BT (3) yields a
(1+1)-dimensional $(N+1)$-component integro-differential system
\begin{eqnarray}
\eqalign{u_y=\sum_{i=1}^N a_i\exp(\int{u-u_i}\rmd x),\cr
u_x+u_{i,x}=-2\lambda_i+\frac{(u-u_i)^2}{2}, \qquad i=1,2,...,N, }
\label{86}
\end{eqnarray}
where $\lambda_i$ and $a_i$ ($i=1,2,...,N$) are constants. \\
By the transformation $u_i=u-(\ln{\phi_{iy}})_x$, equation
(\ref{85}) becomes
\begin{eqnarray}
u=\sum^{N}_{i=1}a_i \phi_i,\label{87}
\end{eqnarray}
which  then converts (\ref{86}) into the following (1+1)-dimensional
N-component differential system
\begin{eqnarray}
2\phi_{ixxy}\phi_{iy}-4\left(\sum^{N}_{m=1}a_m
\phi_{mx}\right)\phi^2_{iy}- \phi^2_{ixy}-4\lambda_i \phi^2_{iy}=0,
 \quad i=1,2,...,N. \label{88}
\end{eqnarray}

Alternatively, combining the  constraint condition (\ref{85}) with
the $t$-part of the BT (4) will produce a (1+2)-dimensional system
about $x$, $y$ and $t$, reading
\begin{eqnarray}
\eqalign{u_y=\sum_{i=1}^N a_i\exp(\int{u-u_i}\rmd x),\cr
u_t+u_{i,t}=2u^2_{x}+2u^2_{i,x}+2u_xu_{i,x}-(u-u_i)(u_{xx}-u_{i,xx}),\quad
i=1,2,...,N. }\label{89}
\end{eqnarray}
Using the same transformation and equation (\ref{88}), equation
(\ref{89})  is transformed into the following  $N$-component system
\begin{eqnarray}
\phi_{ixyt}\phi_{iy}-\phi_{ixy}\phi_{iyt}-2(\sum^{N}_{m=1}a_m\phi_{mxx})\phi_{iy}\phi_{ixy}
+(\sum^{N}_{m=1}a_m\phi_{mx}-2\lambda_i)\phi^2_{ixy} \nonumber\\
+2[(\sum^{N}_{m=1}a_m\phi_{mx})^2+2\lambda_i
\sum^{N}_{m=1}a_m\phi_{mx}-\sum^{N}_{m=1}a_m\phi_{mt}+4\lambda^2_i]\phi^2_{iy}=0.\label{90}
\end{eqnarray}

It should be noted that the integrability of the
infinite-dimensional systems (\ref{88}) and (\ref{90}) obtained in
this way is not quite clear. The finite-dimensional models obtained
here are completely integrable,  that strongly suggests  these
infinite-dimensional models should have many nice integrable
properties. It will be of much interest to investigate the
integrability of these models in the further work.

\section{Conclusion and discussions}

In this paper, we have shown that combining nonlocal symmetries with
BTs can result in many diverse applications.
The main new progresses made in this paper in the general aspect of integrable systems are:\\
(i). The BTs are used to find nonlocal symmetries;\\
(ii). Different types of BTs may possess same infinitesimal forms and
then new types of BTs may be obtained from old ones;\\
(iii). New integrable (negative) hierarchies can be obtained from nonlocal symmetries related to
 BTs;\\
(iv). New finite dimensional integrable systems can be obtained from
BTs and related symmetry constraints and reductions. And then the
original high dimensional model can be solved from lower dimensional
ones because of the existence of nonlocal symmetries depending on BTs ;\\
(v). The exact interaction solutions among solitons and other
complicated waves including periodic cnoidal waves and Painlev\'e
waves are revealed which have not yet found for any integrable
models because it is difficult to solve the original BT (or Darboux
transformation) problem if the original seed solutions are taken as the cnoidal
or Painlev\'e waves;\\
(vi). The localization procedure results in a new way to find Schwartz form of
the original model which is obtained usually via Painlev\'e analysis for the
continuous integrable systems. The method may provide a potential method to transform
a discrete integrable systems to Schwartz forms because usually the BTs of discrete
integrable models are known.

The above progresses are realized especially for potential KdV
(pKdV) equation. For pKdV equation, it possesses a new class of
nonlocal symmetry resulting from its BT. Since this BT is of Riccati
type, more information about its bilinear forms is learned via the
Cole-Hopf transformation. Based on the new nonlocal symmetry with
internal parameters, we construct two sets of negative pKdV
hierarchies and fulfill their corresponding bilinear forms.

In order to  extend applicability of nonlocal symmetry to obtain
explicit solutions of  KdV equation, we introduce another two
auxiliary variables $v$ and $g$ to form a prolonged system with $u$
and $u_1$, so that the original nonlocal symmetry can be transformed
to a Lie point symmetry of the new equations system. Then what
follows naturally are Lie-B\"acklund transformation and two kinds of
novel similarity reductions. By virtue of two kinds of BTs, the
solitary wave solutions of KdV equation are obtained through the
transformations of the trivial solutions. Concerning the complete
Lie point symmetries of the prolonged system, we achieve rich group
invariant solutions including rational solution hierarchy, Bessel
function solution hierarchy and periodic function solutions.

The nonlocal symmetry has also been devoted to construct various new
integrable systems by symmetry constraint method. Applying nonlocal
symmetry on the BT of  pKdV equation, finite-dimensional integrable
systems are given, which are found equivalent to the excellent work
done by Cao \cite{16}. Moreover, the introduction of an internal
parameter as new independent variable helps us to build two sets of
infinite-dimensional models.

We believe that both the negative pKdV hierarchies and  two sets of
infinite-dimensional models obtained in the paper should have many
nice integrable properties. For the completely  integrable
finite-dimensional models, one may consider their algebraic geometry
solutions to achieve related solutions of KdV equation. For the
localization of nonlocal symmetries, it still remains unclear what
kind of nonlocal symmetries must have close prolongations and can be
applied to construct exact solutions. It is quite reasonable that
these matters merit our further study.

\section*{Acknowledgments}
The authors are indebt to thank very much for the referees' comments
and suggestions and the helpful discussions with Profs. X B Hu, Q P
Liu, E G Fan, C W Cao and X Y Tang.
This work is supported by the National Natural Science Foundation of
China (Grant Nos. 11075055, 11175092, 61021004, 10735030), Shanghai
Leading Academic Discipline Project (No. B412) and K C Wang Magna
Fund in Ningbo University.


\section*{References}
\begin{thebibliography}{99}

\bibitem{1} Rogers C and Shadwick W F 1982 {\it{ B\"acklund Transformation and Their
Applications}}
 (New York)
\bibitem{2} Olver P J 1993 {\it {Applications of Lie Groups to Differential Equations}}
(New York: Springer)
\bibitem{3} Bluman G W and Kumei S 1989 {\it{Symmetries and Differential Equations}}
 (Berlin:Springer)
\bibitem{4} Lou S Y 1993 {\it{Phys. Lett. B}} \textbf{302} 261\\
            Lou S Y 1993 {\it{Int. J. Mod. Phys. A}} \textbf{3A} 531\\
            Lou S Y 1994 {\it{J. Math. Phys.}} \textbf{35}  2390
\bibitem{5} Lou S Y 1993 {\it{J. Phy. A: Math. Gen.}} \textbf{26} L789\\
            Lou S Y 1993 {\it{Phys. Lett. A}} \textbf{181} 13\\
            Lou S Y 1994 {\it{Phys. Lett. A}} \textbf{187} 239\\
            Lou S Y 1994 {\it{Solitons, Chaos and Fractals}} \textbf{4} 1961\\
            Ruan H Y and Lou S Y  1993 {\it{J. Phys. Soc. Japan}} \textbf{62} 1917\\
            Han P and Lou S Y 1994 {\it{Acta Phys. Sinica}} \textbf{43} 1042\\
            Lou S Y and Chen W Z 1993 {\it{Phys. Lett. A}} \textbf{179} 271
\bibitem{6} Lou S Y 1997 {\it{J. Phys. A: Math. Phys.}} \textbf{30} 4803
\bibitem{7} Lou S Y and Hu X B 1997 {\it{J. Phys. A: Math. Gen.}} \textbf{30} L95
\bibitem{8} Lou S Y and Hu X B  1997 {\it{J. Math. Phys.}} \textbf{38} 6401
\bibitem{9} Edelen D G 1980 {\it{Isovector Methods for  Equations of Balance}}
            (Alphen aam den Rijn: Sijthoff and Noordhoff)
\bibitem{10} Krasil'shchik I S and Vinogradov A M 1984 {\it{Acta Appl. Math.}}  \textbf{2} 79
\bibitem{11} Krasil'shchik I S and Vinogradov A M  1989  {\it{Aata  Appl. Math.}} \textbf{15} 161
\bibitem{12} Galas F 1992 {\it{J. Phys. A}} \textbf{25} L981
\bibitem{13} Lou S Y and Hu X B 1993 {\it{Chin. Phys. Lett.}} \textbf{10} 577
\bibitem{14} Lou S Y, Ruan H Y, Chen W Z, Wang Z L and Chen L L 1994 {\it{Chin. Phys. Lett.}}
             \textbf{11} 593
\bibitem{15} Cao C W 1990 {\it{Science in china (Series A)}} \textbf{33} 528
\bibitem{16} Cao C W {\it{Acta Mathematica  Sinica}} 1991 \textbf{7} 216
\bibitem{17} Cao C W, Wu Y T and Geng X G 1999 {\it{J. Math. Phys.}} \textbf{40} 3948
\bibitem{18} Antonowicz M and  Rauch-Wojciechowski S 1991 {\it{J. Phys. A}} \textbf{24} 5043
\bibitem{19} Antonowicz M and  Rauch-Wojciechowski S 1992 {\it{J. Math. Phys.}} \textbf{33} 2115
\bibitem{20} Lou S Y 1997 {\it{Commun. Theor. Phys.}} \textbf{27} 249
\bibitem{21} Wahlquist H D and Estabrook F B 1973 {\it{Phys. Rev. Lett.}} \textbf{31} 1386
\bibitem{22} Hu X B, Lou S Y and Qian X M 2009 {\it{Stud. Appl. Math.}} \textbf{122} 305
\bibitem{23} Lou S Y 1998 {\it {Physica Scripta}} \textbf{57} 481
\bibitem{24} Hone1 A N and  Wang J W 2003 {\it{Inverse Problems}} \textbf{19} 129
\bibitem{25} Verosky J M 1991 {\it{J. Math.  Phys.}} \textbf{32} 1733
\bibitem{Gu} Guthrie G A 1993 {\it{J. Phys. A: Math. Gen.}} \textbf{26} L905\\
             Guthrie G A and Hickman M S 1993 {\it{J. Math.  Phys.}} \textbf{26} 193\\
             Guthrie G A 1994 {\it{Phys. R. Soc. Lond. A}} \textbf{446} 107
\bibitem{26} Levi D 1988 {\it{Inverse  Problems}}  \textbf{4} 165
\bibitem{27} Ince E L 1956 {\it{Ordinary Differential Equations}} (New York)
\bibitem{28} Umemura H and Watanabe H 1997 {\it{Nagoya Math. J.}} \textbf{148} 151

\end{thebibliography}
\end{document}